\documentclass[rnote]{aa} 
\usepackage{wasysym}
\usepackage{txfonts}
\usepackage{natbib}

\newcommand{\GYRE}{\texttt{GYRE} }
\newcommand{\CODATA}{\texttt{CODATA} }
\newcommand{\BIS}{\texttt{BiSON} }
\newcommand{\NACRE}{\texttt{NACRE} }
\newcommand{\OPAL}{\texttt{OPAL} }

\usepackage{color}

\begin{document}

\title{A Bayesian estimation of the helioseismic solar age}
\author{A. Bonanno\inst{1}\and
H.-E. Fr\"ohlich\inst{2}
          }
\offprints{Alfio Bonanno\\ \email{alfio.bonanno@inaf.it}}

\institute{
INAF, Osservatorio Astrofisico di Catania, via S. Sofia, 78, 95123 Catania, Italy
\and
Leibniz Institute for Astrophysics Potsdam (AIP), An der Sternwarte 16, 14482 Potsdam, Germany
}
\date{\today}
\abstract
{
The helioseismic determination of the solar age 
has been a subject of several studies because it 
provides us with an independent estimation of the age of the solar system.
}
{
  We present the Bayesian estimates 
of the helioseismic age of the Sun, which are determined by means of  calibrated solar models that employ different 
equations of state and nuclear reaction rates.
}
{
We use 17 frequency separation ratios $r_{02}(n)=(\nu_{n,l=0}-\nu_{n-1,l=2})/(\nu_{n,l=1}-\nu_{n-1,l=1})$ 
from 8640 days of low-$\ell$ \BIS frequencies and consider three likelihood functions 
that depend on the handling of the errors of these $r_{02}(n)$ ratios. 
Moreover, we employ the  2010 \CODATA 
recommended values for Newton’s constant, solar mass, and radius to calibrate a large grid of solar models
spanning a conceivable range of solar ages.
}
{
It is shown that the {most constrained} posterior distribution of the solar age for 
models employing Irwin EOS with \NACRE reaction rates 
leads to $t_\odot = 4.587 \pm 0.007$ Gyr, 
while models employing the Irwin EOS and 
Adelberger, et al., Reviews of Modern Physics, 83, 195 (2011)
reaction rate have $t_\odot = 4.569 \pm 0.006 $ Gyr. Implementing \OPAL EOS
in the solar models results in reduced evidence ratios (Bayes factors) and leads to an age that is
not consistent with the meteoritic dating of the solar system.
}
{
An estimate of the solar age that relies on an helioseismic age indicator such as $r_{02}(n)$ 
turns out to be essentially independent 
of the type of likelihood function.
However, with respect to model selection, abandoning any information concerning 
the errors of the $r_{02}(n)$ ratios leads to inconclusive results, and this  
stresses the importance of evaluating the trustworthiness 
of error estimates. 
}

\keywords{Sun: helioseismology -- Sun: interior -- Sun: oscillations -- Equation of state -- 
          Nuclear reactions, nucleosynthesis, abundances -- Methods: statistical}
\titlerunning{A Bayesian approach to the helioseismic solar age}
\authorrunning{Bonanno}

\maketitle           

\section{Introduction}
By definition, the age of the Sun is the time since the protosun arrived at some reference state
in the HR diagram, usually called the ``birth line'' where it begins its quasi-static contraction \citep{stahler83}.  
For a star of 1 $M_\odot$, this stage depends on various factors, such as rotation, spin, and wind accretion during the protostar phase, 
and it should produce a pre-main-sequence (PMS) object of $\sim 4000$ K with a luminosity of  $\sim 10~L_\odot$  \citep{stapa}.

The age of the solar system is instead assumed to lie between the age of 
crystallized and melted material in the solar system, and the time of 
significant injection of nucleosynthesis material in the protosolar nebula.
The latter is an upper limit to the solar age, and the precise 
relation between the ``age of the Sun'' and that of the planetary bodies and nebular
ingredients depends on the detailed physical conditions during the collapse of the 
protosolar nebula. 

Recent studies based on the dating of calcium-aluminum-rich inclusions (CAI) in chondrites
have reported a  meteoritic ``zero age'' of the solar system ranging 
roughly from 4.563 to 4.576 Gyr \citep{ba95,amelin02,jacobsen08,jacobsen09,bouvier10}. 
A more recent estimate supports an age of 4.567  Gyr \citep{conn}. 
The interesting question is to locate this time in the PMS evolution of the Sun as defined

To what extent tis he solar system age consistent with the notion of ``birth line'' and
its location in the HR diagram?
In fact, the PMS evolution is often neglected in stellar evolution calculation, 
based on the duration of this phase being much shorter than the successive evolution. On the other hand,
the precise location of the zero-age main-sequence (ZAMS) is problematic, as discussed in 
\cite{morel00}.

Several studies  have thus tried to use helioseismology to provide an independent estimation of the solar age, thus  
testing the consistency of the radioactive dating  of the solar system 
with stellar evolution theory.
The standard approach to  calibrating helioseismic solar age
has been to confronting specific oscillations diagnostic among solar models of different ages,
while keeping the luminosity, the radius, and the heavy elements abundance at the surface fixed
\citep{douglas90,dz99,age02,joergen09}. 
The resulting helioseismic age is then a ``best-fit'' age 
that can be obtained with a series of calibrations.
The limitations of the one-parameter calibration approach \citep{douglas01,houdek11} lie in the difficulties of estimating the 
effect of chemical composition and, in general for unknown physics \citep{bomu}, of determining 
the sound speed gradient near the center \citep{dough12}. 

The essential ingredient for estimating the helioseismic solar age is to find a ``genuine''  
oscillation diagnostic for which the sensitivity to the complex physics of the outer layers of the star 
is minimal.  
As is well known \citep{rox03,oti05}, the frequency separation ratio 
$r_{l,l+2}(n) = (\nu_{n,l}-\nu_{n-1,l+2})/(\nu_{n,l+1}-\nu_{n-1,l+1})$ of low-degree p-modes 
is a fairly good indicator of the inner structure of the sun because it is  mostly sensitive 
to the gradient of mean molecular weight near the center.
In particular in \cite{boda}, it was shown that $r_{02}(n)$ is a relatively robust age indicator since it
does not depend on the surface-effect 
correction of the higher order p-modes \citep{kb08} or on different definitions of the 
solar radius. 

In this work we discuss a Bayesian approach to the helioseismic determination
of the solar age. In fact in recent times, the use of the Bayesian inference has become increasingly common in the astronomical community 
\citep{trotta}. In the case of asteroseismology, the use of Bayesian inference
has been essential for establishing credible and robust intervals for parameter estimation of stellar  and solar modeling \citep{seek,bazo,gru13}.
In  \cite{grug} it has been argued that
that there is an inconsistency between the meteoritic and the solar age as inferred from helioseismology.
In our case the advantage of using the Bayesian inference as opposed to the standard frequentist's approach lies in  
the possibility 
of rigorously comparing the relative plausibility of solar models 
with different physical inputs, chemical compositions, etc., by computing evidence ratios (i.e., Bayes factors). 

Besides comparing models with different EOS and nuclear reaction rates,  we use 
the new recommended 2010 \CODATA values for the solar mass, the Newton constant, 
and the solar radius, and we check the robustness of our findings  by using different likelihood functions to quantify the 
agreement of model predictions with the \BIS data \citep{broom}.
We show that updated physical inputs
lead to an helioseismic age that is consistent with the meteoritic age of the solar system. 
{as also suggested from the analysis of \cite{grug}.}
On the other hand, at the level of accuracy of {current} helioseismology, it is essential to include
the PMS evolution (about 40--50 Myr) in the standard solar model calibration.  

The structure of this work is the following. 
Section 2 compiles the physical details of the solar models considered and the observations, 
Sect. 3 describes the Bayesian approach, Sect. 4
presents the results, and Sect. 5 is devoted to the conclusions. 

\section{Models physics and observations} 
Our non-rotating solar models were built with the Catania version of the GARSTEC code
\citep{garstec}, a fully-implicit 1D code that includes heavy-elements diffusion. 
It employs either the \OPAL 2005 equation of state \citep{rogers96,rogers02}, complemented
with the MHD equation of state at low temperatures \citep{hummer}, 
or the Irwin equation of state \citep{cassisi},
and it uses \OPAL opacities for high temperatures \citep{iglesias}
and Ferguson's opacities for low temperatures \citep{ferguson}.
The nuclear reaction rates are either from  the \NACRE collaboration \citep{angulo} 
or from the \cite{adelberger} compilation, and the chemical 
composition follows the mixture of \cite{gn93} with 
$(Z/X)_\odot=0.0245$ at the surface. 
We also consider models with the so-called ``new abundances'' 
for which  $(Z/X)_\odot=0.0178$ \citep{asplund09}.

Our starting models are chemically homogeneous PMS models with $\log L/L_\odot = 0.21$ and
$\log T_e = 3.638 $ K, so they are fairly close to the birth line of a $1 M_\odot$ object, 
{which according to \cite{stapa}, would be located  about 2 Myr before}. 
The value of Newton's constant is the 2010 \CODATA recommended value
$G=6.67384 \times 10^{-8}$ ${\rm cm^3 g^{-1}s^{-2}}$
taken from \url{http://physics.nist.gov/cuu/Constants/index.html}.
As a consequence, because $ G M_\odot \equiv \kappa = 1.32712440 \times 10^{26}$ ${\rm cm^3 s^{-2}}$
\citep{cox},  we assumed $M_\odot = 1.98855 \times 10^{33} \rm g$. 
The radius is taken to be $R_\odot = 6.95613 \times 10^{10} \rm cm$
based on an average of the two values and quoted error bar in Table
3 of \cite{haber}. (See also \cite{reese} for an application of the 2010 \CODATA 
to seismic inversions.) The solar luminosity is instead 
$L_\odot = 3.846 \times 10^{33}~\rm erg~s^{-1}$ \citep{cox}.

We then used the definitive ``best possible estimate''  
of  8640 days of low-$\ell$ frequency \BIS data, corrected for the
solar cycle modulation \citep{broom} 
taken from \url{http://bison.ph.bham.ac.uk/index.php?page=bison,frequencies}.
In particular, we considered $N = 17$ frequency separation ratios 
$r_{02}(n)=(\nu_{n,l=0}-\nu_{n-1,l=2})/(\nu_{n,l=1}-\nu_{n-1,l=1})$, 
ranging from order $n=9$ to order $n=25$ for the $l=0,1,2$ 
modes, together with the corresponding uncertainties. 

\section{Bayesian inference}
In the Bayesian view the central quantity to be computed is a conditional 
probability $P(A|B)$: 
It represents the probability that event $A$ will happen
given the fact event $B$ has actually occurred. In our case it is   
a {\it \emph{subjective}\/} type of probability, a degree of plausibility of $A$ given $B$,
and as such it bears no resemblance to a frequency distribution.
In common applications $A$ is a vector of parameters that quantify a model 
and $B$ the set of observational data.  Estimates of the parameters
have the same conceptual status as probabilistic events. 

According to Bayes' theorem, 
\begin{equation}
P(A|B) = P(A) \frac{P(B|A)}{P(B)}
\label{bayes}
\end{equation}
holds, where $P(A)$ and $P(B)$ are unconditional probabilities for events $A$ and $B$.
In our application $P(A|B)$ is the probability that model parameter $\tau$ falls within the 
interval $\tau\dots{}\tau+\mathrm{d}\tau$ given the data. $P(B|A)$ is the usual 
likelihood function, $P(A)$ the so-called prior for $\tau$, and $P(B)$ the searched-for 
{\it \emph{evidence}\/} of the model. The evidence is a model's {\it \emph{mean}\/} likelihood 
 -- averaged over the whole $\tau$ range and subject to 
the prior probability density -- and as such, it measures the overall ability of 
the model to cope with the data. 
Since the prior probability must sum up to unity, the evidence diminishes if 
the parameter space expands beyond the space needed to cover the essential parts of 
the likelihood mountain.

We are interested here in evidence ratio, the Bayes factors. They 
allow a model ranking, i.e. to compare the explanatory powers of competitive 
models or hypotheses. The logarithm of age, $\tau = \log_\mathrm{e}(t)$, is chosen, with $t$ being the
solar age, 
to make certain that the posterior for the age is compatible with 
the posterior of, say, the reciprocal of the age.
With this decision the evidence does not depend on the unit of age. 
Accordingly, for $P(A)$ a flat prior is assumed over the logarithm of age. 
All the other parameters of the solar model -- the initial helium 
fraction $Y_0$, the mixing length parameter, and the initial $(Z/X)_0$ value -- 
are adjusted in order to reproduce measured radius, effective temperature,
and $Z/X$ ratio at the surface. We do not attempt to derive credibility intervals
for $Y_0$, $(Z/X)_0$, and the mixing length parameter, because they are obtained {\it \emph{analytically}}, i.e., by means of the standard Newton-Raphson procedure embedded
in the solar model calibration. 

In actual calculations the likelihood function $P(B|A)$ is often expressed by a Gaussian. 
In particular, if $d_i=r_{02}(n)$ are the observed data ($n=i+8, i=1\dots{}N$), 
$m_i$ the theoretical model values, and $\sigma_i$  the errors, it reads
as\begin{equation}
\Lambda(\tau)=\prod_{i=1}^{N}\frac{1}{\sqrt{2\pi}\sigma_i}\exp\left(-\frac{(d_i-m_i(\tau))^2}{2\sigma_i^2}\right)\,,
\label{l1}
\end{equation}
with $\tau$ being $\log_\mathrm{e}(t)$ and $N=17$. 
An observed quantity, $d_i = r_{02}(i+8)$, is the ratio of two Gaussians that results, 
strictly speaking, in a Cauchy-like distribution\footnote{One can indeed substitute Eq.~(\ref{l1}) by the correct expression. The
corresponding ages and their standard deviations are indistinguishable from
the values communicated in Table~\ref{tab:Gauss}. However, as the distribution is
a ``fat-tailed'' one, all Bayes factors but one (that normalized to unity)
are somewhat enhanced: in the worst case (``Irwin+AdelR+Asplund'') by 44 per cent,
in all other cases by up to 14 per cent. More important, the ranking is not
affected.}, not a Gaussian. With respect to the following likelihood function, which is
based on Eq.~(\ref{l1}), {and} for the sake of consistency, we decided, {however,} to compute
the $\sigma_i$'s according to Gauss's linearized error propagation rule. 

Although \cite{broom} have carefully taken the systematic error 
induced by solar activity into account, it is tempting to enhance the robustness of the age estimates 
by averaging over the errors, thereby obeying Jeffreys' $1/\sigma$ prior.
In this way an error-integrated likelihood is obtained, which relies still on the Gaussianity assumption:
\begin{equation}
\Lambda^\prime =
\int_0^\infty \hspace{-5pt}\Lambda\,{{\rm d}\sigma\over\sigma} =
{\Gamma\left(N/2\right)\over 2\,\left(\pi N\right)^{N/2}}{1\over\left[\prod_{i=1}^{N} s_i\right]
\left[{1\over N}\sum_{i=1}^N\left({d_i-m_i}\over s_i\right)^2\right]^{N/2}}\,, \label{l2}
\end{equation}
where $\sigma_i = s_i\cdot\sigma$, and relative errors $s_i$ are normalized according to
$\sum_{i=1}^N{w_i} = N$, {with $w_i = 1/s_i^2$ the weights}.

We also consider the median likelihood, 
which is constructed from the binomial distribution and ignores any 
information with regard to measurements errors. 
It only assumes that both positive and 
negative deviations $d_i-m_i$ have equal probability. 
It reads
as\begin{equation}
\Lambda^{\prime\prime} =
\prod_{i=1}^{N}\frac{1}{2^N}\frac{N!}{K!(N-K)!}\,, \label{l3}
\end{equation}
with $K$  the number of events with $d_i\geq m_i$ (or $d_i\leq m_i$).

The evidence is the {\it \emph{mean}\/} likelihood over parameter space; that is to say, in the case of a flat prior for $\tau= \log_\mathrm{e}(t)$, 
\begin{equation}
\frac{1}{\tau_\mathrm{u}-\tau_\mathrm{l}}\int_{\tau_\mathrm{l}}^{\tau_\mathrm{u}}
\Lambda^{(n)} \,\mathrm{d}\tau\,,
\end{equation}
where the interval $\tau_\mathrm{u}-\tau_\mathrm{l}$ is the same in all 
cases where, when considered otherwise, the communicated Bayes factors would render useless.

\begin{table}
\caption{Expectation value and standard deviation of the solar age 
for the Gaussian likelihood function (\ref{l1}) with known errors $\sigma_i$. 
The ordering of the models is according to their Bayes factors.} 
\centering                         
\begin{tabular}{l r l }       
\hline\hline
\\[-8pt]         
Input physics &  $t_\odot\mathrm{[Gyr]}\;\;\;\;\;$  & Bayes factor  \\    
\hline
\\[-8pt]
  Irwin\,+\NACRE           &$4.587\pm 0.007$&$1                 $ \\[2pt]
  Irwin\,+AdelR            &$4.569\pm 0.006$&$0.17              $ \\[2pt]
  \OPAL+\NACRE+old \CODATA &$4.696\pm 0.006$&$2.1\times 10^{-13}$ \\[2pt]
  \OPAL+\NACRE             &$4.702\pm 0.006$&$1.9\times 10^{-15}$ \\[2pt]
  \OPAL+AdelR              &$4.683\pm 0.008$&$3.0\times 10^{-17}$ \\[2pt]
  Irwin\,+AdelR+Asplund    &$4.785\pm 0.006$&$2.9\times 10^{-44}$ \\[2pt]
\hline                                
\hline                                
\end{tabular}
\label{tab:Gauss}
\end{table}

\begin{table}
\caption{Ages and Bayes factors 
in the case of unknown magnitude of the over-all error $\sigma$ in Eq.~(\ref{l2}).} 
\centering                         
\begin{tabular}{l r l }       
\hline\hline
\\[-8pt]         
Input physics &  $t_\odot\mathrm{[Gyr]}\;\;\;\;\;$  & Bayes factors\\    
\hline
\\[-8pt]
  Irwin\,+\NACRE       & $4.587\pm 0.010$  & $1                $ \\[2pt]
  Irwin\,+AdelR        & $4.569\pm 0.010$  & $0.42             $ \\[2pt]
  \OPAL+\NACRE+old \CODATA & $         4.697\pm 0.017$ & $2.3\times 10^{ -4}$ \\[2pt]
  \OPAL+\NACRE         & $4.703\pm 0.018$  & $1.0\times 10^{-4}$ \\[2pt]
  \OPAL+AdelR          & $4.683\pm 0.018$  & $5.6\times 10^{-5}$ \\[2pt]
  Irwin\,+AdelR+Asplund& $4.786\pm 0.026$ & $1.1\times 10^{ -7}$ \\[2pt]
\hline                                
\hline                                
\end{tabular}
\label{tab:Jeff}
\end{table}

\begin{table}
\caption{Ages and Bayes factors in the case of median statistics, Eq.~(\ref{l3}). 
The ordering of the models is as in Tables \ref{tab:Gauss} and \ref{tab:Jeff}.}  
\centering                         
\begin{tabular}{l r l }       
\hline\hline
\\[-8pt]         
Input physics &  $t_\odot\mathrm{[Gyr]}\;\;\;\;\;$  & Bayes factors\\    
\hline
\\[-8pt]
  Irwin\,+\NACRE       & $4.582\pm 0.014$  & $0.38$ \\[2pt]
  Irwin\,+AdelR        & $4.563\pm 0.015$  & $0.43$ \\[2pt]
  \OPAL+\NACRE+old \CODATA &         $ 4.692\pm 0.026$ & $0.60$ \\[2pt]
  \OPAL+\NACRE         & $4.698\pm 0.028$  & $0.63$ \\[2pt]
  \OPAL+AdelR          & $4.676\pm 0.029$  & $0.65$ \\[2pt]
  Irwin\,+AdelR+Asplund& $4.775\pm 0.039$  & $1$    \\[2pt]
\hline                                
\hline                                
\end{tabular}
\label{tab:Binom}
\end{table}

\section{Results}

To properly resolve $\Lambda(\tau),$ the $\tau$ domain was covered by 80 equidistant grid points. 
We checked that our results are substantially insensitive to a further refinement of the grid. 
(The achieved log time resolution is, if expressed in a musician's language, a little bit better than one cent.)
The computation of the eigenfrequencies has been performed with the latest version of the public pulsation code,  \GYRE
\url{https://bitbucket.org/rhdtownsend/gyre/wiki/Home}, which employs a new Magnus multiple-shooting 
scheme, as described in detail in \cite{gyre}. For consistency we have thus implemented the 2010 \CODATA 
also in \GYRE.
The results, ages, and Bayes factors are depicted in Tables~\ref{tab:Gauss}--\ref{tab:Binom} for 
the three likelihood functions (\ref{l1}--\ref{l3}) considered. 

First, we note that the use of a significantly longer data set \citep{broom},
in combination with a Gaussian likelihood (\ref{l1}), has provided us with a much sharper solar age estimate
if compared to previous studies \citep{boda,joergen09}; 
models obtained with Irwin EOS generally perform better than models with OPAL EOS,
while the evidence for models with \NACRE and \cite{adelberger} reaction rates
differ only marginally. Only \cite{adelberger}
reaction rates, combined with an Irwin EOS result in a helioseismic age  $t_\odot$, are consistent with the meteoritic one 
within $1\sigma$.  

The main reason for the age difference between models with \NACRE and \cite{adelberger}
reaction rates is the value of the astrophysical $S_\mathrm{pp}(0)$-factor for the pp-fusion
cross-section.  The  \NACRE collaboration adopts $S_\mathrm{pp}=3.89 \times 10^{-25}$ MeVb,
while \cite{adelberger} adopt $S_\mathrm{pp}=4.01\times 10^{-25}$ MeVb. As a consequence,
the helioseismic age tends to be longer for the \NACRE rates because
of the reduced efficiency of the pp reactions. 

Using the 2010 \CODATA does not produce a significant impact 
on the solar age determination. On the other hand,  
if we consider for the \OPAL+\NACRE 
case in Table~(\ref{tab:Gauss}), which is the old \CODATA value for Newton's constant
$G=6.67232 \times 10^{-8}$ ${\rm cm^3 g^{-1}s^{-2}}$,
we find  $t_\odot = 4.696 \pm 0.006$ Gyr and a Bayes factor 
that is two orders of magnitude larger.
This value for the solar age  is consistent within
one $1\sigma$ with the value of $t_\odot = 4.62 \pm 0.08$ Gyr found by 
\cite{boda} (including PMS evolution), which was  obtained with  
a much shorter \BIS data set. 

In the case of the error-integrated likelihood (\ref{l2}), 
the expectation values of the solar age are reported in Table~\ref{tab:Jeff}.
They are basically the same
as in the former (Gaussian) case (Table~\ref{tab:Gauss}), but the uncertainties 
are {at least} a factor two larger. Of course, this is due to the 
extreme stance that even the magnitude of the error $\sigma$ in 
(\ref{l2}) is assumed to be unknown. 

Even more robust ages estimates are presented 
in Table~\ref{tab:Binom}. In median statistics, 
nothing is assumed about the shape of the {\it \emph{symmetric}\/} error distribution.  
The expectation values of the solar age are 
systematically lower by up to 10 Myr as compared with the the 
two other cases where the Gaussianity of the 
errors is presumed. 
The credibility regions are up to seven times larger than in the case of 
a Gaussian likelihood with trusted errors. 
More important, in terms of Bayes factors, all models now perform equally well!

\section{Conclusions}
{ Bayes factors are powerful indicators when it comes to quantifying the ability of solar models
differing in input physics to cope with published ``quiet Sun'' frequency separation ratios.
If one trusts the common Gaussianity assumption, models using the Irwin EOS perform best.
Abandoning any error information, to be on the safe side, leads to inconclusive results with
respect to Bayes factors (cf. Table~\ref{tab:Binom}),
which stresses the importance of evaluating the trustworthiness of the error estimates that enter into 
Eqs.~(\ref{l1} and \ref{l2}). 


If we assume a Gaussian error distrobution, only models with Irwin EOS agree with the meteoritic age. Moreover, their Bayesian evidence
exceeds those with an OPAL EOS by at least four orders of magnitude (cf. Table~\ref{tab:Jeff}).
With the Adelberger et al. reaction rates, the solar age proves to be even more consistent with the meteoritic age.

Incorporation of PMS evolution is essential because the standard deviation of the age estimation,
$\le 10$ Myr in the case of Irwin EOS, is less than the 40--50 Myr time span of the PMS phase.
Moreover, our helioseismic age is consistent with the notion of a birth line, since our starting PMS
models are only 2 Myr distant from the birth line location.
The switch from old CODATA to 2010 CODATA values does not effect the age estimate significantly, in contrast to the
model's evidence, at least if the published errors are taken seriously. }

These conclusions are obtained using the ``old'' 
\cite{gn93} with 
$(Z/X)_\odot=0.0245$ at the surface. 
It is worth mentioning that the  so-called ``new abundances','
for which  $(Z/X)_\odot=0.0178$ \citep{asplund09},
would lead to completely inconsistent values of the solar age and 
Bayes factors reduced by many orders of magnitude as indicated by 
the corresponding row entries in Tables \ref{tab:Gauss} and \ref{tab:Jeff}. 

\vspace{0.5cm}
{\it Acknowledgments}.  AB would like to thank Gianni Strazzulla and Brandon Johnson for
enlightening discussions about calcium-aluminum-rich inclusions 
and the meteoritic age of the solar system.

\end{document}